\documentclass[fleqn,usenatbib]{mnras}
\usepackage{newtxtext,newtxmath}
\usepackage[T1]{fontenc}
\usepackage{ae,aecompl}

\usepackage{multirow}
\usepackage{comment}
\usepackage{pdflscape}
\usepackage{graphicx}
\usepackage{colortbl}
\usepackage{array}
\newcolumntype{C}[1]{>{\centering\let\newline\\\arraybackslash\hspace{0pt}}p{#1}}
\usepackage{siunitx}
\usepackage{longtable}
\usepackage{tabu}
\usepackage{enumerate}

\newcommand{\Ion}[2]{#1{\,\small #2}}
\newcommand{\Teff}{\mbox{$T_\mathrm{eff}$}}
\newcommand{\logg}{\mbox{$\log g$}}
\title{NGTS and HST insights into the long period modulation in GW\,Librae}
\author[P.~Chote et al.]{
\parbox{\textwidth}{
P.~Chote$^{1}$\thanks{E-mail: p.chote@warwick.ac.uk},
B.\,T.~G\"ansicke$^{1}$,
J.~McCormac$^{1}$,
A.~Aungwerojwit$^{2}$,
D.~Bayliss$^{1}$,
M.\,R.~Burleigh$^{3}$,
S.\,L.~Casewell$^{3}$,
Ph.\,Eigm\"uller$^{4}$,
S.~Gill$^{1}$,
M.\,R.~Goad$^{3}$,
J.\,J.~Hermes$^{5}$,
J.\,S.~Jenkins$^{6,7}$,
A.\,S.~Mukadam$^{8}$,
S.~Poshyachinda$^{9}$,
L.~Raynard$^{3}$,
D.\,E.~Reichart$^{10}$,
P.~Szkody$^{8}$,
O.~Toloza$^{1}$,
R.\,G.~West$^{1}$,
P.~J.~Wheatley$^{1}$
}
\vspace{0.1cm}
\\
$^1$Department of Physics, University of Warwick, Coventry CV4 7AL, United Kingdom\\
$^2$Department of Physics, Faculty of Science, Naresuan University, Phitsanulok 65000, Thailand\\
$^3$Department of Physics and Astronomy, University of Leicester, Leicester, LE1 7RH, UK\\
$^4$Institute of Planetary Research, German Aerospace Center, Rutherfordstrasse 2, 12489 Berlin, Germany\\
$^5$Department of Astronomy, Boston University, Boston, MA 02215, USA\\
$^6$Departamento de Astronom\'ia, Universidad de Chile, Camino El Observatorio 1515, Las Condes, Santiago, Chile\\
$^7$Centro de Astrof\'isica y Tecnolog\'ias Afines (CATA), Casilla 36-D, Santiago, Chile\\
$^8$Department of Astronomy, University of Washington, Box 351580, Seattle, WA 98195, USA\\
$^9$National Astronomical Research Institute of Thailand (Public Organization), Chiangmai, 50180, Thailand\\
$^{10}$Department of Physics and Astronomy, University of North Carolina at Chapel Hill, Chapel Hill, NC 27599, USA\\
\vspace{-0.9cm}
}

\begin{document}

\date{}
\pubyear{}

\pagerange{\pageref{firstpage}--\pageref{lastpage}}
\maketitle

\label{firstpage}

\begin{abstract}

Light curves of the accreting white dwarf pulsator GW\,Librae spanning a 7.5 month period in 2017 were obtained as part of the \textit{Next Generation Transit Survey}. This data set comprises 787 hours of photometry from 148 clear nights, allowing the behaviour of the long (hours) and short period (20\,min) modulation signals to be tracked from night to night over a much longer observing baseline than has been previously achieved. The long period modulations intermittently detected in previous observations of GW\,Lib are found to be a persistent feature, evolving between states with periods $\simeq$ 83\,min and 2 -- 4\,h on time-scales of several days. The 20\,min signal is found to have a broadly stable amplitude and frequency for the duration of the campaign, but the previously noted phase instability is confirmed. Ultraviolet observations obtained with the Cosmic Origin Spectrograph onboard the \textit{Hubble Space Telescope} constrain the ultraviolet-to-optical flux ratio to $\simeq 5$ for the 4\,h modulation, and $\lesssim 1$ for the 20 minute period, with caveats introduced by non-simultaneous observations. These results add further observational evidence that these enigmatic signals must originate from the white dwarf, highlighting our continued gap in theoretical understanding of the mechanisms that drive them.

\end{abstract}


\section{Introduction}

It has now been over 20 years since coherent short-term variability was discovered in the dwarf nova GW\,Librae \citep{Warner1998}, and attributed to non-radial pulsations of the central white dwarf. This revealed a new class of accreting white dwarf pulsators, of which more than a dozen are now known: all residing in short-period, low accretion rate, cataclysmic variables (CVs) \citep[e.g.][]{Warner1998, Woudt2004, Warner2004, Araujo-Betancor2005, Vanlandingham2005, Patterson2005, Gaensicke2006, Nilsson2006, Mukadam2007, Patterson2008, Pavlenko2009, Woudt2011, Uthas2012, Mukadam2017}. Their physical and asteroseismological properties differ noticeably from those of the well-studied single white dwarf pulsators with hydrogen-envelopes (DAV, ZZ\,Ceti stars, \citealt{Mukadam2004, Gianninas2016, VanGrootel2012}) and helium-envelopes (DBV, V777\,Her stars, \citealt{Beauchamp1999}): the envelopes of accreting white dwarf pulsators are hydrogen-dominated, but enriched by non-negligible amounts of helium and trace metals from their companion star. The H/He ratio affects the driving of pulsations, and has hence to be considered as a third dimension, effectively establishing an instability volume rather than the two-dimensional strips in \Teff\ and \logg. Consequently, an additional HeII partial ionisation zone may form that can drive pulsations at higher temperatures compared to the ZZ\,Ceti stars \citep{Townsley2004, Arras2006, VanGrootel2015}. In fact, the known accreting white dwarf pulsators span a wide range of effective temperatures from 10500\,K to above 15000\,K \citep{Szkody2010}. In addition, white dwarfs in CVs can be spun up to very short periods \citep[e.g.][]{King1991,Cheng1997}; $\sim$100\,s for GW Lib \citep{vanSpaandonk2010}, compared with $\sim$ tens of hours for typical isolated white dwarfs \citep[e.g.][]{Hermes2017}, which further complicates interpretation of the observed pulsation signals. 

Observations of GW\,Lib obtained between 1997 and 2005 determined a binary period of 76.78 minutes \citep{Thorstensen2002} from H$\alpha$ radial velocity measurements, and established the presence of three pulsation modes near 650, 380, and 230\,s visible in both optical and ultraviolet light curves \citep{vanZyl2000, Szkody2002, vanZyl2004, Copperwheat2009}. While these three pulsation signals were consistently detected, they did not have the frequency stability seen in many ZZ\,Ceti or V777\,Her white dwarf pulsators, and the spread of periods observed between runs and the apparent splitting of frequencies within individual long runs was interpreted as signs of true underlying structure \citep{vanZyl2004}. In addition to these short-period pulsation modes, \citet{Woudt2002} discovered a quasi-periodic 2.1\,h modulation, whose origin remained unexplained.

The CVs hosting accreting white dwarf pulsators undergo dwarf novae `superoutbursts' with recurrence times of tens of years, during which the accretion disc reaches a critical density, becomes unstable, and rapidly dumps its contents onto the surface of the white dwarf. This material compresses and heats the envelope \citep{Godon2003, Piro2005} so that its temperature increases to a value outside the instability strip, halting pulsations. Monitoring the white dwarf as it returns to its quiescent temperature provides an opportunity to study the onset and evolution of pulsations over human time-scales -- a few years, as compared with $\sim10^8\,$years for evolutionary cooling. 

GW\,Lib was discovered on 1983 August 10, when it underwent a superoutburst \citep{Maza1983}, i.e. the white dwarf pulsations were identified after the system had been 14 years back into quiescence. The next superoutburst was detected on 2007 April 11 \citep{Templeton2007}, and observations during the outburst only revealed superhumps from the accretion disc (\citealt{Vican2011}, hereafter VI11). The first short-period modulations were detected in 2008, though not at the periods seen prior to the outburst, but at $\simeq19$\,min and $\simeq290$\,s (\citealt{Copperwheat2009}, hereafter CO09), in line with the expectations that heating of the envelope during the superoutburst affects the driving of the pulsations. 
GW\,Lib was intensively monitored at optical and ultraviolet wavelengths over the next decade, and presented overall a consistent behaviour, exhibiting brightness modulations at $\simeq20$\,min and $\simeq300$\,sec (\citealt[hereafter SC10]{Schwieterman2010}; VI11; \citealt[hereafter BU11]{Bullock2011}; \citealt[hereafter SZ12]{Szkody2012}; \citealt[hereafter CH16]{Chote2016}; \citealt[hereafter TO16]{Toloza2016}; \citealt[hereafter SZ16]{Szkody2016}; \citealt[hereafter CH17]{Chote2017}). Two puzzling facts emerged from this vast array of data: (1) After an initial brief phase of cooling (SZ12), the white dwarf has remained at a constant temperature since 2011, and is still $\simeq3000$\,K hotter than in quiescence (TO16, SZ16)~--~contrasting observations of similar large amplitude outburst dwarf novae which showed the white dwarfs monotonously cooling back to their pre-outburst temperature on a time-scale of $\simeq5$\,yr, e.g. WZ\,Sge \citep{Slevinsky1999, Godon2006} and AL\,Com \citep{Szkody2003}. (2) The long-period modulation seen before the outburst was initially detected at 2.1\,h in 2008 in both the optical (CO09) and UV (BU11), but now often shows up at $\simeq4$\,h, which led VI11 to suggest that this is the fundamental of the $\simeq2.1$\,h signal and that this modulation originates in the accretion disc. However, TO16 demonstrated that the $\simeq4$\,h brightening event captured by ultraviolet spectroscopy was clearly related to the heating and cooling of the white dwarf in GW\,Lib. A more detailed assessment of the long-period modulations remains particularly difficult as their time-scale is similar to that of a typical observing run.

GW\,Lib was observed in the survey footprint of the Next Generation Transit Survey (NGTS, \citealt{Wheatley2018}) between January and September 2017. Here we report on the analysis of 787\,h of time-series photometry obtained over 148 clear nights spanning 7.5 months. The observations and data reduction are described in \S\ref{section:obs}, before analysing the long- and short-timescale variability in \S\ref{section:longp} and \S\ref{section:shortp}. Finally, \S\ref{section:discussion} attempts to put these results into context against earlier observations of GW\,Lib and other systems before closing in \S\ref{section:conclusions} with our conclusions.

\section{Observations and Data Reduction}\label{section:obs}
\subsection{NGTS}

The NGTS facility \citep{Wheatley2018} consists of 12 robotic 20\,cm telescopes that are housed in a common enclosure at the European Southern Observatory (ESO) Cerro Paranal site in Chile. Each telescope observes a $2.8\times2.8$ degree field of view through a broad red (roughly $R+I$) filter using a $2\text{k}\times2\text{k}$ back-illuminated deep-depleted CCD camera, producing a $5$\,arcsec per pixel plate scale.

The NGTS survey strategy, as it operated in 2017, observed pre-selected fields continuously with the same telescope every night while $\gtrsim 30^{\circ}$ elevation. Closed-loop auto-guiding \citep{McCormac2013} ensured that stars were observed using the same CCD pixels for the entire observing season. The standard survey exposure was 10\,s and it took 3\,s to read out the CCD, giving a survey cadence of 13\,s.

The survey field NG1518-2518 containing GW\,Lib was scheduled for observations between 2017 January 25 and 2017 September 21, but we chose to truncate the data set on 2017 September 15 when the observability dropped below 1\,h each night. This provided a data set containing 218045 exposures from 148 nights, corresponding to a total of 606\,h integration time from 787\,h time on target.

Due to issues with the shutter, the camera was unmounted twice during the season. When the camera is remounted there is a small shift in the field rotation and hence a new auto-guiding reference image is acquired for each subset of nights.

GW\,Lib is too faint to be included in the standard NGTS analysis pipeline, so we developed a stand-alone reduction procedure that operated on the raw images. This allowed us to automatically extract light curves at the end of each night and follow the behaviour of GW\,Lib in near-real time. The final reduction procedure operates as follows:

\begin{figure*}
\centering
\includegraphics[angle=90,width=\textwidth]{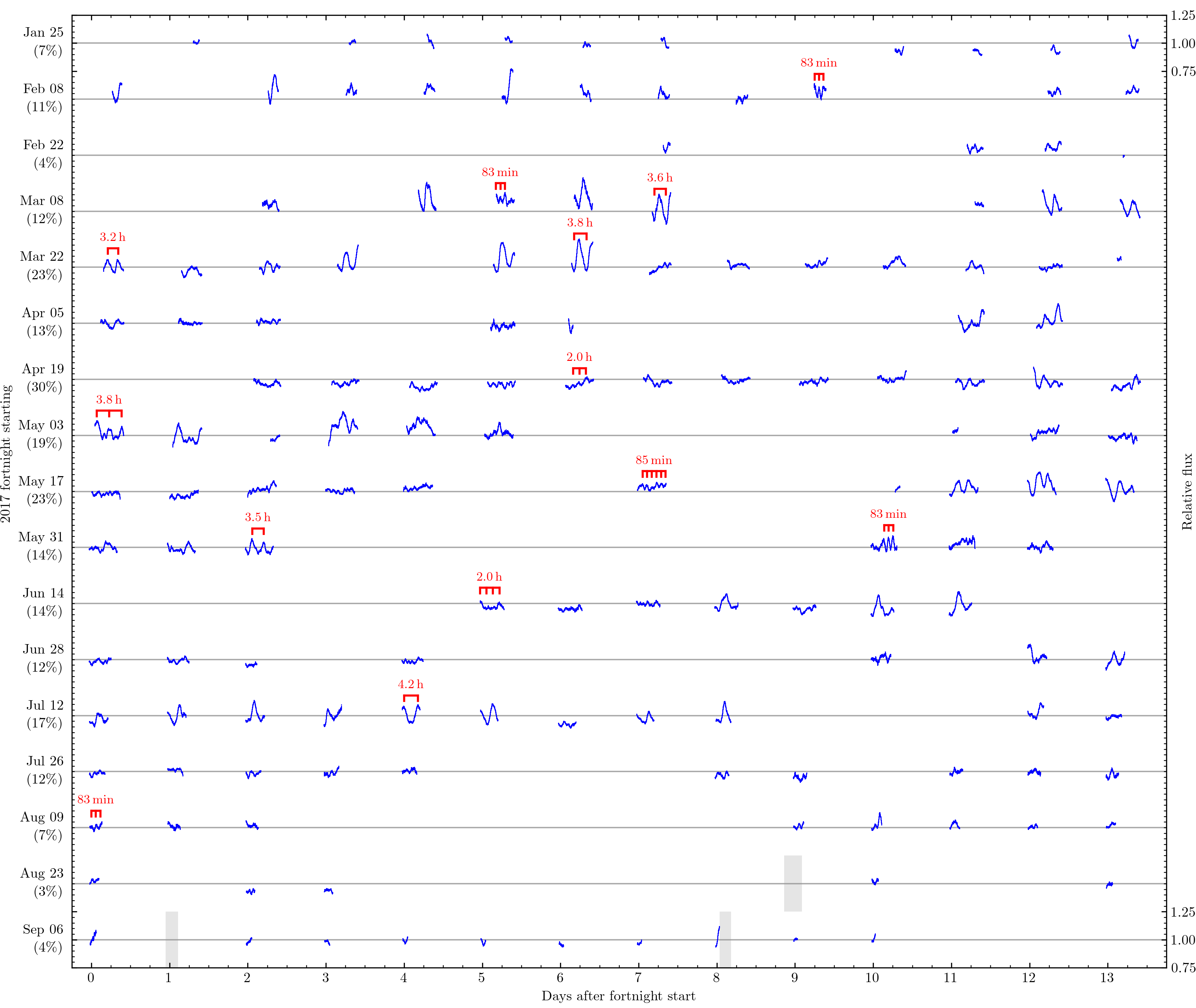}
\caption{The full 7.5 month NGTS light curve of GW\,Lib, smoothed with a 40\,min boxcar filter to emphasize the long-period behaviour. Each fortnight is offset vertically by 0.5 units, and the observation duty cycle for each fortnight is listed in parentheses below the start date. The regular gaps in the light curve are due to the day/night cycle; the gaps on February 1 -- 2 and August 11 -- 15 were due to technical issues; the remaining gaps are nights when NGTS was not operating due to cloud. Several nights are marked with indicative periods to highlight the detection of the previously observed long period modulations. The time windows of the \textit{HST} observations are indicated with shaded rectangles.}
\label{figure:longperiod-full}
\end{figure*}

\begin{enumerate}[(1)]
  \item A deep-stack reference frame is manually created for each auto-guider reference by averaging 100 frames from the same night's data.
  \item Accurate pixel positions are determined in each reference frame for GW\,Lib and five nearby comparison stars. Comparison stars were selected based on their brightness and lack of blending with nearby stars and confirmed to be non-variable by inspection of their light curves.
  \item A SExtractor \citep{Bertin1996} segmentation map is generated using the SEP Python library \citep{Barbary2016}. The segmentation map is converted to a pixel mask that identifies pixels that are suitable for calculating the background sky level.
  \item Master bias and sky flat frames are created for each night using the afternoon and morning calibration frames. On nights where calibration frames were not taken or when the sky flats were affected by cloud (verified by manual inspection) they are substituted with frames from the closest good night.
  \item Science frames are prepared by (optionally) applying the calibration frames and then subtracting a background sky map that is calculated using the standard SEP/SExtractor routine over the pixels defined by the reference background mask. 
  \item Aperture positions for the comparison stars are calculated by adding the auto-guider offset (specified in the frame metadata) to the reference positions. The apertures are then recentred to correct for the small drifts that occur each night as the field distortion changes with airmass. The aperture for GW\,Lib is placed using a blind offset from the mean comparison star position.
  \item Aperture photometry is done using one of two ways:
  \begin{enumerate}
    \item Photometry for the long period analysis in Section~\ref{section:longp} is extracted using the SEP circular aperture routines with a 2.5\,pixel aperture radius. This method minimises systematic effects at the expense of increased noise in a given measurement. This essentially white noise bins down as we combine many measurements using a running mean.
    \item A second set of photometry is extracted omitting the bias and flat field corrections and using a smaller 1.5\,pixel aperture radius for GW\,Lib while keeping the 2.5\,pixel radius for the comparison stars. This extraction is used in Section \ref{section:shortp} where it improves the sensitivity to the short time-scale variability at a trade-off of introducing long ($\simeq$\,hours) time-scale systematic trends.
  \end{enumerate}  
  \item The extracted flux measurements and UTC timestamps are converted to tabular files containing UTC and BJD$_\text{TDB}$ time, relative flux, and relative flux error using the \textsc{tsreduce} \citep{Chote2014} software. These files, along with the other data described below, are made available as supplementary data in the online journal.
\end{enumerate}

As GW\,Lib is bluer than most of the nearby stars, differential extinction leads to reduced flux ratios at larger airmasses. This effect is reduced in the NGTS data due to the redder bandpass and limited airmass range of observations, but it can still be measured and removed following the technique from CH16. The target / comparison flux ratio is fitted with a linear coefficient as a function of airmass simultaneously for the entire data set. The phase and amplitude of the long period modulations average out over many nights, allowing the airmass effect to be extracted much more robustly than the standard technique of fitting a low-order polynomial to each run.

\subsection{APO \& Prompt}

Additional observations were obtained on 2017 March 4 using the Agile photometer on the 3.5\,m telescope at Apache Point Observatory, and on 2017 May 5 and 2017 May 6 using the 60\,cm robotic Prompt8 telescope at CTIO. The APO observations consisted of 515 30\,s integrations obtained through a BG39 filter, and the Prompt observations consisted of 253 and 300 30\,s integrations obtained through a Baader Clear (C) filter.

The APO and Prompt data were reduced following our standard procedures \citep[see e.g.][SZ16]{Chote2014} to extract aperture photometry, with one exception: the polynomial fit to remove long-period trends was omitted as there was insufficient data to separate the effect of differential extinction from the long-period variability intrinsic to GW\,Lib.

\subsection{\textit{Hubble Space Telescope}}
Quasi-simultaneous \textit{Hubble Space Telescope (HST)} far-ultraviolet (UV) spectroscopy of GW\,Lib was obtained using the Cosmic Origins Spectrograph (COS, \citealt{Green2012}) on the nights of 2017 August 31, September 6, and September 13, near the end of the NGTS observation window. The first two observations occurred during bad weather at Paranal, so simultaneous observations were not possible. The weather was clear during the final run, but GW\,Lib set past the NGTS elevation limit less than 2\,min before the \textit{HST} observations started.

The time-tag COS data were reduced following the procedure from TO16 to create ultraviolet light curves. Data were binned into 5\,s sub-spectra, which were then background-subtracted and the Ly$\alpha$ and O\Ion{I}~airglow lines at 1216 and 1302\,\AA\ were masked. Each spectrum was then binned in the spectral direction to create individual photometric points with a 5\,s cadence. We present limited results from this data set where they provide important context to the NGTS observations, but defer the full analysis to a future publication (Toloza et al, in prep).

\section{Long Period Variability}\label{section:longp}

Photometry demonstrating the long-period behaviour was extracted using a 2.5 pixel aperture and all calibrations. Points more than 5 $\sigma$ from the nightly mean (42 in total) were rejected, and then the data were smoothed with a running 40\,min mean (roughly 190 points) to reduce the scatter introduced by noise and the 20\,min signal. The resulting 34-week long light curve is presented in Fig. \,\ref{figure:longperiod-full}.

The familiar signals near 83\,minutes and between 2 -- 4\,h from earlier observations (illustrated in CH17) are clearly visible, and we can see for the first time these signals switching on and off, with each state lasting for \mbox{1 -- 2} weeks. The larger amplitude modulations appear to grow rather haphazardly out of the quieter states.

\begin{figure}
\centering
\includegraphics[width=\columnwidth]{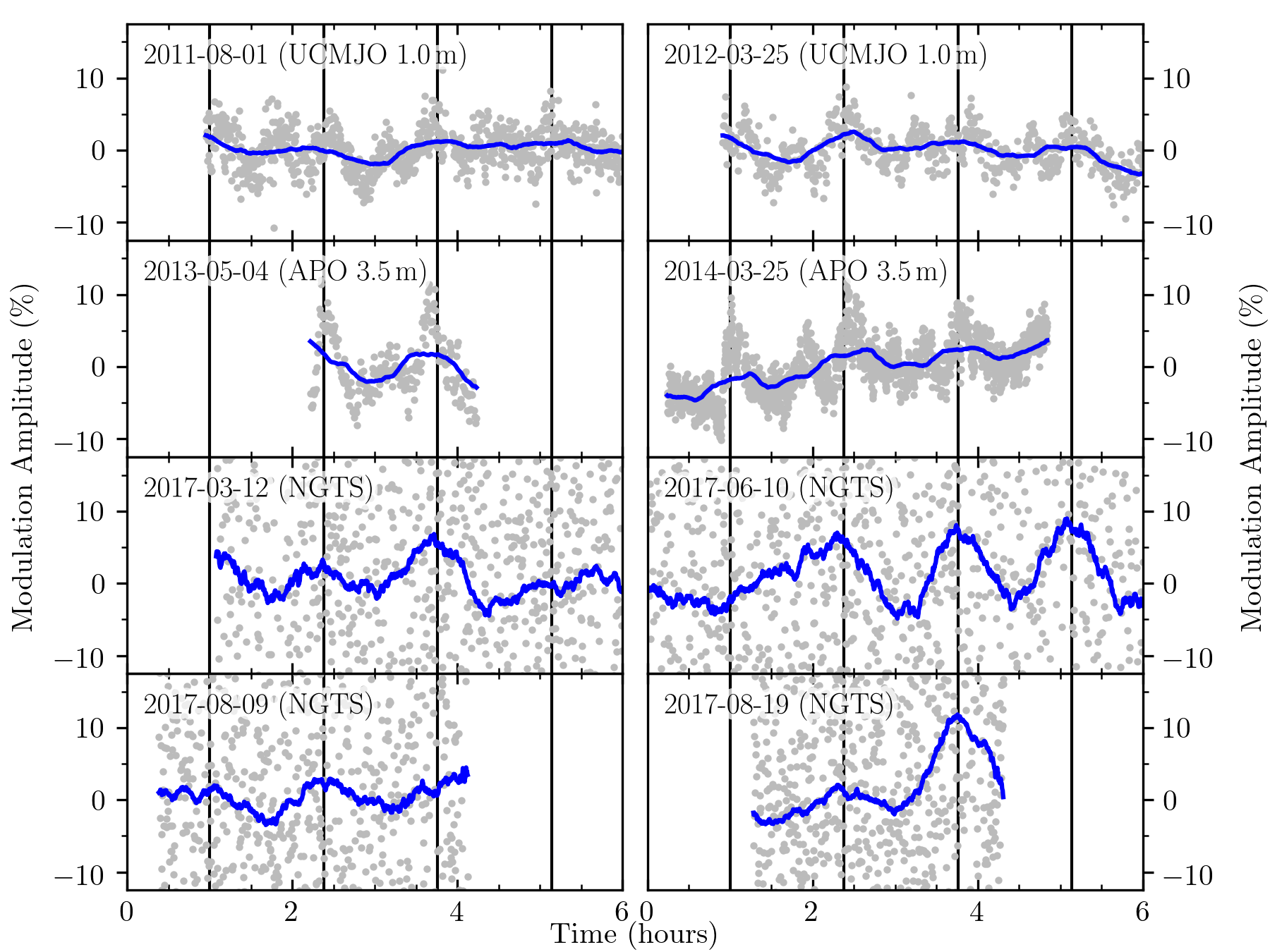}
\caption{The 83\,min periodicity has been seen frequently over the last several years with a ragged multi-peaked appearance. The top panels show examples observed with larger telescopes (from SZ12, CH16, CH17), and the bottom panels show examples from the NGTS data set. Vertical black lines denote the common 83\,min period, and the blue curve shows the  running 40\,min mean.}\label{figure:longp-83min}
\end{figure}

Fig. \ref{figure:longp-83min} compares the 83\,min signals seen on several nights of NGTS observations with earlier observations from larger aperture telescopes. The high-cadence archival observations demonstrate the ragged nature of this signal, which is smoothed out by the running mean. Similar average profiles are visible in the NGTS light curves, and so we infer that this behaviour was present but not resolved due to the low signal-to-noise ratio in the raw NGTS photometry. A few nights in the smoothed NGTS data show amplitudes in excess of 10\%, which implies potentially much larger changes in the unresolved variability, more than has been seen in earlier observations.

\begin{figure}
\centering
\includegraphics[width=\columnwidth]{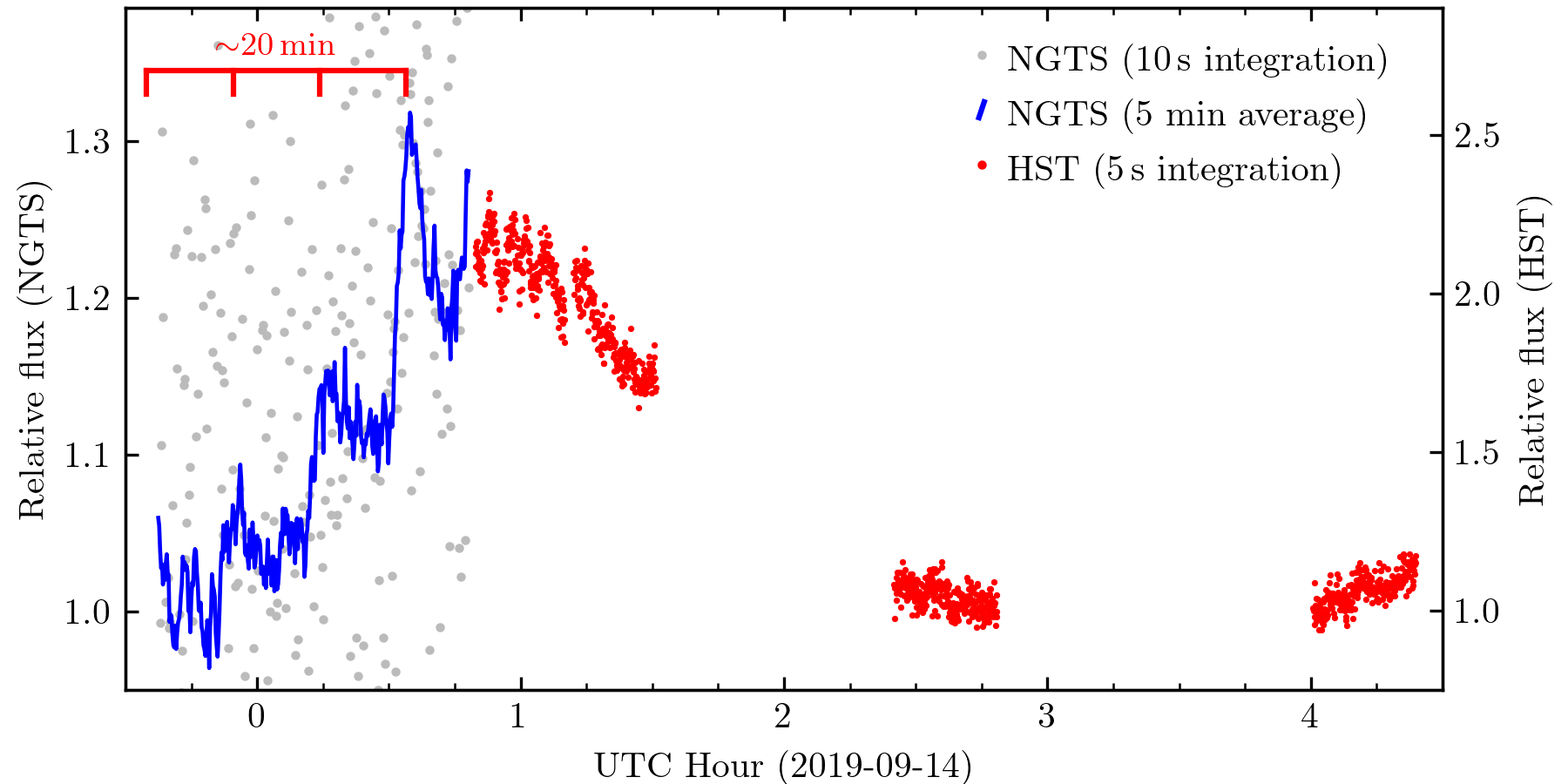}
\caption{Near-simultaneous observations from NGTS and HST on the night of 2017 September 13 captured a cycle of the $\simeq4$\,h modulation, confirming that these modulations are visible in the UV. The disparity in UV/Optical mode amplitudes for the 20\,min and 300\,s periods is also visible. Note that different scales have been used for the NGTS vs \textit{HST} flux.}\label{figure:hst2017c}
\end{figure}

Fig. \ref{figure:hst2017c} presents the near-simultaneous optical and UV light curves on 2017 September 13. The NGTS light curve captures the rise on one of the long period modulations, and the HST light curve captures the subsequent fall.

The full NGTS light curve shows that the brightness in successive minima of the long period modulation is relatively constant (at the 10--20\% level), so we can estimate that the UV to optical amplitude ratio, $A_{\text{UV}} / A_{\text{Optical}} \simeq 1.15 / 0.23 \simeq 5$ assuming that the same holds true in the UV, and that the modulation in the UV and optical are in phase. This is comparable to the UV/Optical ratios measured for the pulsation periods \citep[e.g.][SZ16]{Szkody2002}, reinforcing the suggestion from TO16 that this long-period variability arises from the white dwarf rather than the accretion disc. 

\section{Short-period signals}\label{section:shortp}

To improve the sensitivity to the short-period pulsations we extracted photometry omitting the bias and flat field corrections, and using a smaller (1.5 pixel) extraction radius for GW\,Lib. Reducing the aperture size removes the sky and read noise contributed by the pixels in the wings of the PSF where the signal is negligible. Omitting the calibration frames avoids their noise contribution, taking advantage of the precise sub-pixel telescope guiding to know that any per-pixel effects will remain consistent between exposures. The stars do move on the CCD due to field distortion changes with airmass, but the systematic trends this introduces are on a $\sim$hours time-scale, much longer than the periods of interest (200 -- 2000 seconds).

The time series data is converted to the frequency domain by normalising light curves relative to their mean level and calculating the Discrete Fourier Transform (DFT). Results are presented as amplitude spectra where the vertical scale corresponds to the amplitude of an equivalent sinusoid in the time domain data. The window function for each night is computed by taking the DFT of a noise-free sinusoid sampled at the same times as the observations, and shows the pattern of aliases that are associated with each `real' sinusoidal signal in the time domain data. 

\begin{figure}
\centering
\includegraphics[width=\columnwidth]{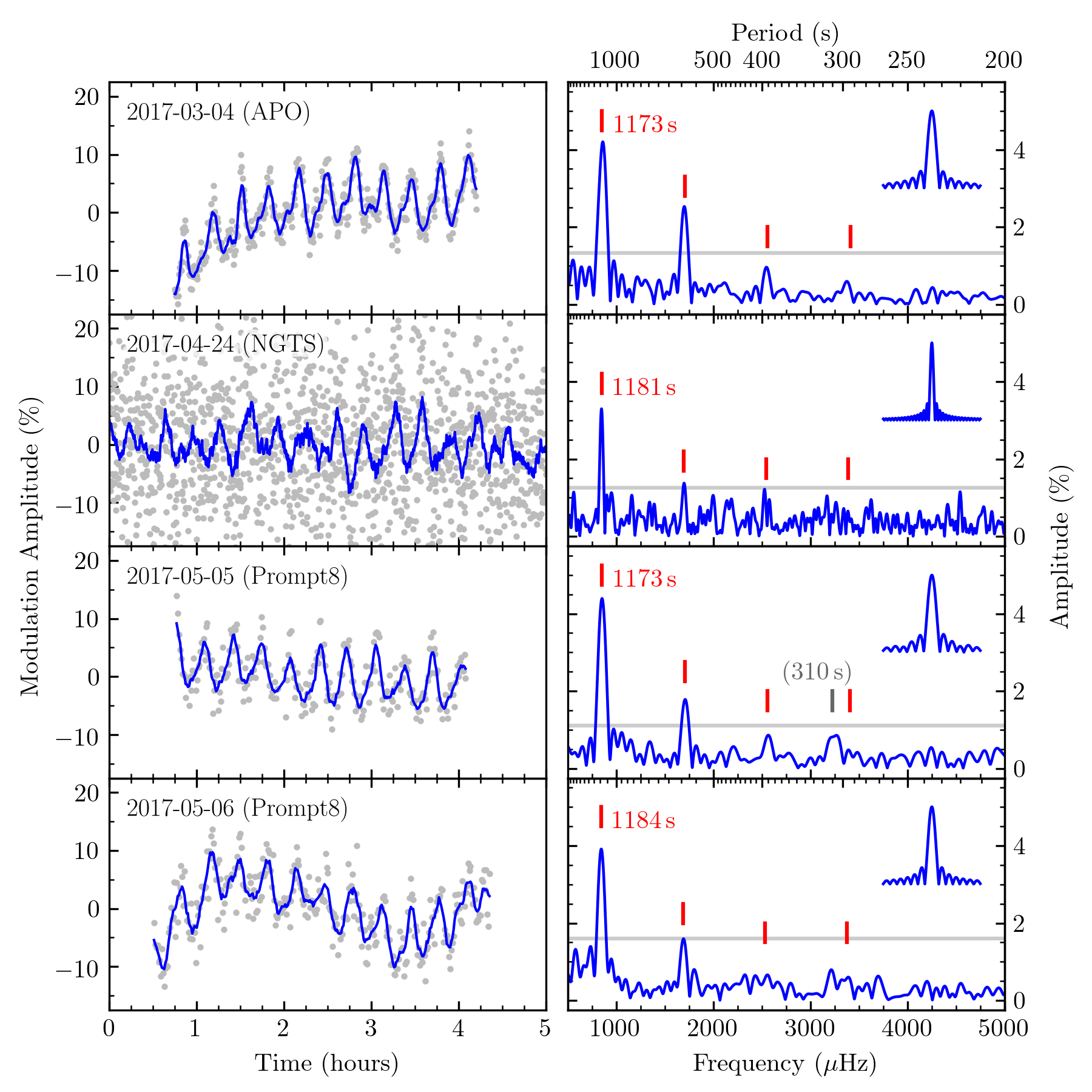}
\caption{Light curves and DFTs of the APO and Prompt8 runs, with a segment of one NGTS run for comparison. The left column presents raw photometric measurements as grey dots, with a blue line showing a running 5\,min mean. The $\simeq4$\,h modulation is clearly visible in the 2017 May 6 run. The right column presents the DFTs and window functions for each night, with the 20\,min period and its harmonics marked with vertical red dashes and the 0.1\% FAP significance thresholds shown as horizontal grey lines. The vertical grey dash on 2017 May 5 indicates a potential but non-significant detection of a 310\,s pulsation period.}\label{figure:runcomparison}
\end{figure}

Fig. \ref{figure:runcomparison} presents an illustrative NGTS light curve alongside the APO and Prompt light curves obtained during the NGTS campaign. The 20\,min period is clearly visible as the dominant feature in the light curves, and the NGTS DFTs compare surprisingly well to the larger aperture telescopes in a large part due to the much longer run lengths (up to 8.8\,h per night).

\begin{figure*}
\centering
\includegraphics[width=\textwidth]{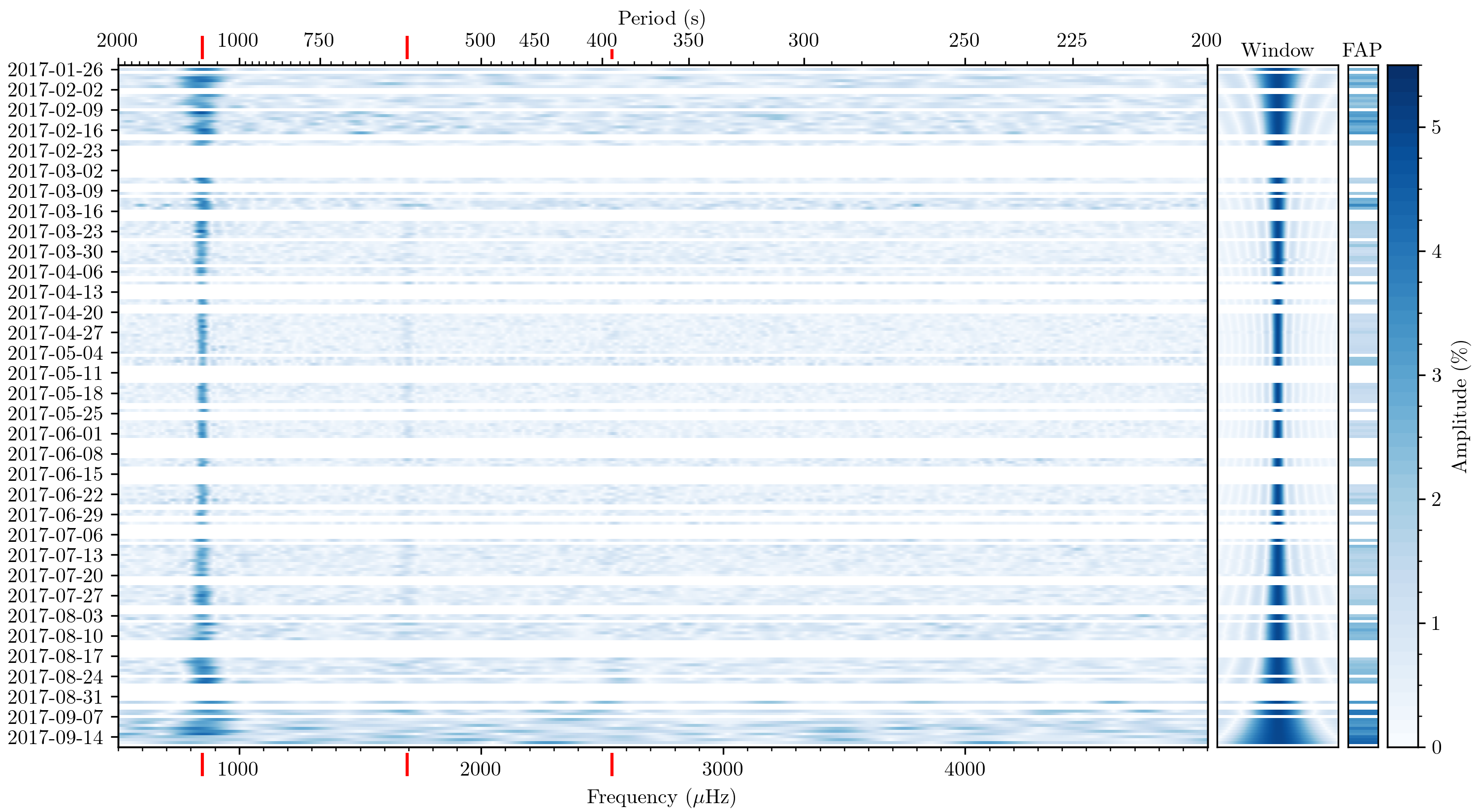}
\caption{Trailed amplitude spectra computed from the NGTS light curves. Each row represents the colour-coded amplitude of the DFT. The Window column shows the associated window function for each night using the same frequency scale. The FAP column indicates the amplitude at which a feature in the DFT is deemed to be statistically significant. The vertical red dashes indicate the location of the 20\,min period and its harmonics. Periods longer than 2000\,s are omitted as they are contaminated by systematic effects introduced by the reduction procedure.}\label{figure:20mintrail}
\end{figure*}

We define an amplitude threshold for statistical significance by applying a False Alarm Probability \citep[FAP; e.g.][]{Sullivan2008} test. The 20\,min signal and its first harmonic are subtracted from the run, and any residual coherent signals are destroyed by randomly shuffling the times associated with each measurement. The DFT is calculated covering 200~--~2000\,s, and the amplitude of the largest peak is recorded. This is repeated 1000 times to simulate different realizations with the same white noise characteristics. The largest peak then defines the amplitude where there is a 1 in 1000 (0.1\%) probability that a peak of the same amplitude can be produced by white noise.

Fig.~\ref{figure:20mintrail} presents the nightly runs as a single trailed DFT. Each horizontal slice shows the DFT, window function, and FAP threshold of the corresponding night with amplitude represented using colour. Several partial nights that were interrupted by bad weather (ten in total) and have a very poor window function were excluded to aid readability.

The 20\,min period is clearly visible at $\simeq$1200\,s with a $\simeq3\,\%$ amplitude throughout the entire 7.5 month observing campaign. Traces of the first and second harmonics can be seen by eye near 590 and 390\,s. The $\simeq300\,$s periods that have been repeatedly seen in the far-ultraviolet (SZ12, SZ16) and occasionally in the optical (CO09, SZ12, SZ16, TO16, CH16) are not detected in the NGTS data. The historical amplitude of this signal was typically $\simeq 0.5 - 2$\% in the optical, and so they would only be apparent above the $\simeq0.7$\% mean noise floor if they remained coherent over several nights -- previous observations (e.g. SZ16, CH17) have shown this is usually not the case.

\begin{figure*}
\centering
\includegraphics[width=\textwidth]{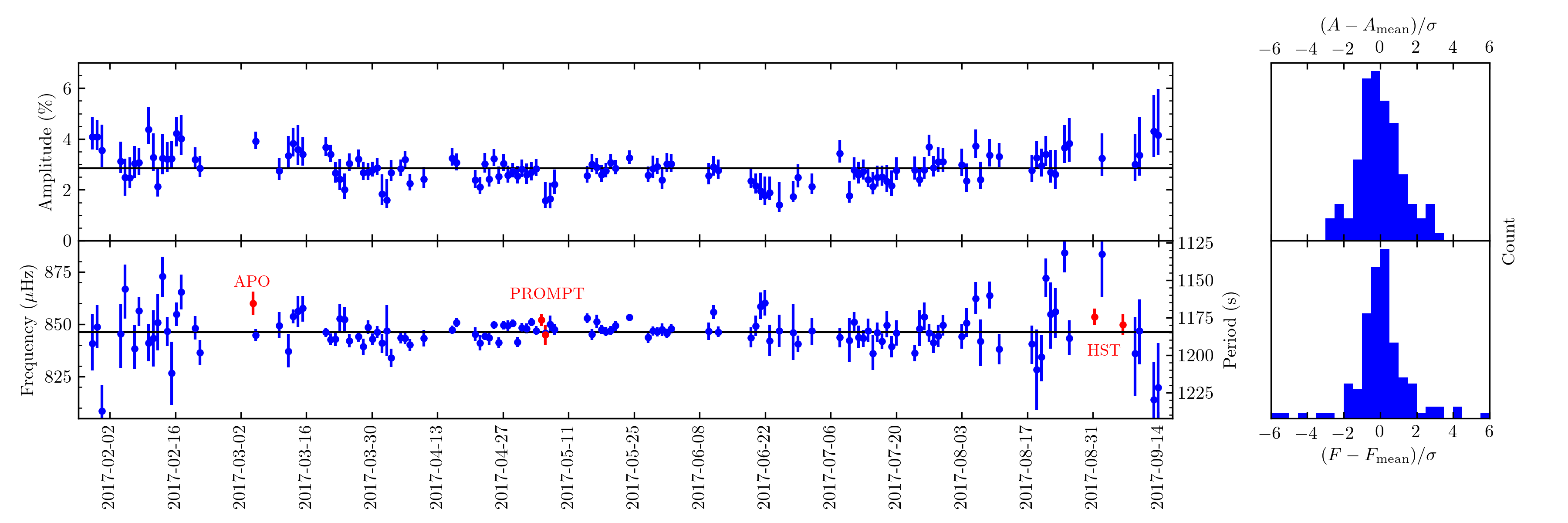}
\caption{Amplitude and frequency/period fits to the 20\,min signal in the NGTS run (blue) and APO, Prompt, and \textit{HST} (see Section \ref{section:longp}) runs (red). The histograms on the right show that the majority of measurements are within $3\sigma$ of the mean 1182\,s period and 2.9\% amplitude.}\label{figure:20minfit} 
\end{figure*}

The period and amplitude of the 20\,min signal on each night were fitted with a sinusoid in the time domain. $1\sigma$ uncertainties were bootstrapped by resampling the light curves with replacement \citep{Efron1979} to generate 5000 alternative realizations of each light curve, which were each fitted to produce a distribution of frequency and amplitude values. The majority of these showed the expected normal distribution, which were fitted with a Gaussian to measure $\sigma$. Nights with $\sigma > 20\,\mu$Hz (five in total) or non-Gaussian profiles (a further nine) were excluded as they do not robustly measure a unique fit.

Fig. \ref{figure:20minfit} presents the fitted frequency/period and amplitudes, which show that the 20\,min signal remains broadly stable in both frequency and amplitude across the full run. The phase and frequency instability shown in CH16 remains, however, as DFTs calculated over multiple nights were found to smear power in the frequency domain in a similar way.

\begin{figure}
\centering
\includegraphics[width=\columnwidth]{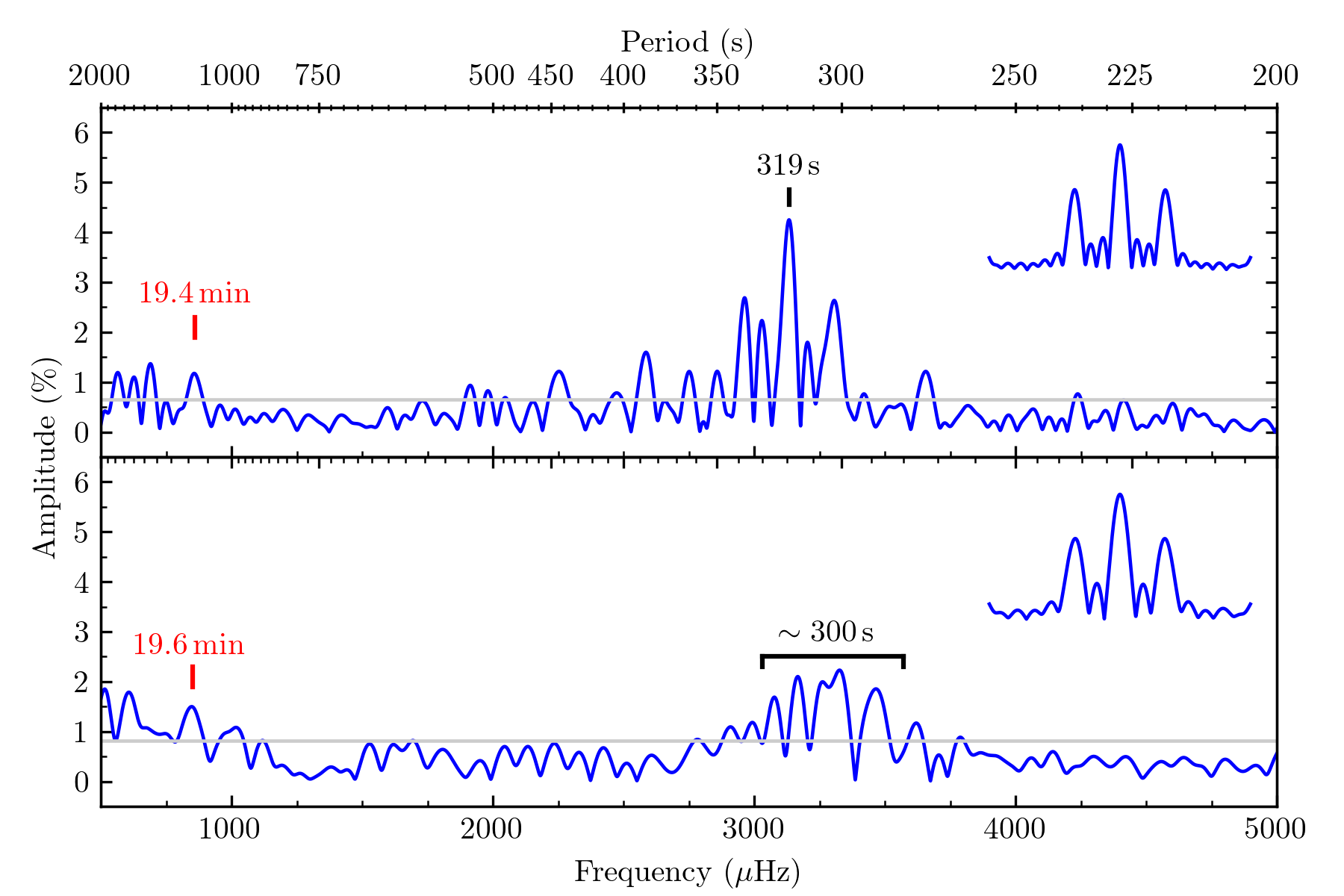}
\caption{DFTs of the 2017-08-31 and 2017-09-06 HST COS observations show that the $\sim$20\,min signal is only weakly detected at periods consistent with the optical observations. The $\sim300\,$s pulsation period dominates the observed variability, but demonstrates poor coherancy in the second visit with power being smeared over a range of frequencies in the DFT. }\label{figure:hst2017ab}
\end{figure}

Fourier transforms of the \textit{HST}/COS observations obtained on 2017 August 31 and 2017 September 6 are presented in Fig. \ref{figure:hst2017ab}. Both observations were dominated by the 300\,s period, which had an amplitude in the range of 2 -- 4\%. Previous observations (SZ12, SZ16) have shown $A_{\text{UV}} / A_{\text{Optical}} \approx 5$, which implies an optical amplitude below 1\% -- consistent with the NGTS non-detection. A more detailed analysis on the behaviour of this pulsation mode will be presented in a future publication.

The key result from the perspective of the NGTS observations is that the 20\,min signal is only weakly detected in the UV, with an amplitude < 2\% constraining $A_{\text{UV}} / A_{\text{Optical}} \lesssim 1$. These observations were obtained during a gap in NGTS observations, so we cannot rule out that the modulation was intrinsically weaker at the time of the \textit{HST} observations, but this seems unlikely considering that the six months of optical measurements before the observation and the few nights after show no evidence of significant amplitude variability.

\section{Discussion}\label{section:discussion}

The 83\,min signal was originally thought to be associated with a late superhump period (BU11). As it has now been regularly observed to come and go for nearly 10 years, it is likely related to the `quiescent superhump' that has been seen to come and go in EQ\,Lyn \citep{Mukadam2013} and V455 And \citep{Araujo-Betancor2005}. In EQ Lyn, the absence of any period in the UV while near-simultaneous optical data showed superhumps suggested that the periodicity may come from the outer accretion disc.

The UV flux from GW\,Lib is dominated by the white dwarf, with the cooler accretion disc contributing just a few percent (see e.g. SZ12) to the total. The accretion disc is more significant at optical wavelengths (VI11 suggest up to 40\% of the total flux), diluting the amplitude of any signals from the white dwarf. Our near-simultaneous measurements showing the amplitude of the $\simeq4\,$h signal as $\simeq5$ times larger in the UV than the optical therefore further supports the suggestion from TO16 that this signal originates from the white dwarf. The GALEX data from BU11 showed that this signal had a larger amplitude in FUV than NUV, pointing to the inner disc or white dwarf, but the optical amplitude was also comparable. Unfortunate timing of the inter-orbit gaps in the 2015 observations (SZ16) mean that we can neither definitively detect nor rule out the presence of the 2\,h signal in the UV at that time.

The NGTS observations appear to show that the behaviour switches betweenthese 2 -- 4\,h and 83\,min states on a time-scale of roughly a week, which suggests that they may share a common source.

The 20\,min signal is similarly enigmatic. It was not recognized as a persistent feature in the pre-2007 outburst photometry, despite \cite{vanZyl2004} noting a detection in their Table 2. There have been at least three extended periods in 2008, 2012, and 2015 -- 2017 where this signal has been the dominant source of variability. It was initially suggested (BU11, VI11) that this signal was a disc-related feature due to its poor phase coherency between nights and its prior non-detections. In the years since, it has become clear that it must originate on the WD, as there are no known mechanisms that could generate signals that maintain the level of observed stability over such a long time-scale.

\cite{Hermes2017} showed that the stable pulsations in isolated DAVs also lose phase coherency at periods greater than 800\,s, so the lack of coherency of the 20\,min signal does not exclude the pulsation mode hypothesis.

The UV/optical amplitude ratio $A_{\text{UV}} / A_{\text{Optical}} < 1$ is more problematic, as the current understanding of g-mode pulsations predict that the pulsation amplitude should increase significantly in the UV \citep{Robinson1995} due to increased limb darkening reducing the geometric cancellation effect. It is not unprecedented, however: \cite{Szkody2010} report similar ratios for two other accreting pulsators (the 1285\,s period in PQ And and the 582 and 655\,s periods in REJ 1255+266); \cite{Kotak2004} shows $A_{\text{UV}} / A_{\text{Optical}} \simeq 1$ for the 272 and 304\,s modes in the prototypical `stable' DAV G117-B15A; and \cite{Kepler2000} likewise for the 141\,s mode in G185-32. So, while we do not yet understand the origin of these signals, the amplitude ratios alone are not sufficient to rule out an origin on the white dwarf.

\cite{Saio2019} presents a case for the variability in these accreting systems not being g-mode pulsations at all, but rather r-mode oscillations trapped in the H-rich layers of the WD. While they can model the shorter period modulations in GW Lib, they are not able to explain the $\sim$4 hour modulations and do not attempt to model the 20 minute signal.

\section{Conclusions}\label{section:conclusions}

Observations of GW Librae between 2017 January 25 and 2017 September 21 provide the longest continuous monitoring campaign yet obtained on this accreting white dwarf pulsator.

These observations demonstrate that the long-period modulations in the system appear to change between states on a time-scale of days/weeks. While the individual modes are not strictly periodic, they do appear to have distinctive and repeating characteristics that can be identified in archival observations going back nearly ten years. Near-simultaneous NGTS and \textit{HST} observations confirm that the 4\,h modulation seen by (TO16) in the UV has the same origin as in the optical, and provide an estimate of the UV/Optical flux ratio of $\simeq 5$.

The steady presence of the 20\,min signal over the full observation baseline strongly suggests that this signal originates on the WD, and is not a transient disc phenomenon. The weak detection of this signal in UV is difficult to explain, but has been seen before in both accreting and non-accreting WD pulsators.

These NGTS observations add significantly to the body of observational data on GW Lib, which document several phenomenological behaviours that so far remain unexplained. Future advancements in explaining these observations will require theoretical developments towards the pulsation (or other potential) mechanisms that can drive near-coherent variability in these systems.

\section*{Acknowledgments}
Based on data collected under the NGTS project at the ESO La Silla  Paranal Observatory. The NGTS facility is operated by the consortium institutes with support from the UK Science and Technology Facilities Council (STFC) under projects ST/M001962/1 and ST/S002642/1. The research leading to these results has received funding from the European Research Council under the European Union's Seventh Framework Programme (FP/2007-2013) / ERC Grant Agreement n. 320964 (WDTracer). BG, OT and PJW received support from the UK STFC consolidated grants ST/L000733/1 and ST/P000495/1. PS and AM acknowledge support from HST GO-13807 and GO-14912.002-A, as well as NSF AST-1514737. AA received support from the Program Management Unit for Human Resources \& Institutional Development, Research and Innovation grant B05F630110. JSJ acknowledges support by FONDECYT grant 1201371 and partial support from CONICYT project Basal AFB-170002. This work is based on observations made with the NASA/ESA \textit{Hubble Space Telescope}, obtained at the Space Telescope Science Institute, which is operated by the Association of Universities for Research in Astronomy, Inc., under NASA contract NAS 5-26555. These observations are associated
with program \#14912. This work has made use of observations obtained with the Apache Point Observatory (APO) 3.5-meter telescope, which is owned and operated by the Astrophysical Research Consortium (ARC). This work has made use of data obtained by the PROMPT-8 telescope, owned by National Astronomical Research Institute of Thailand, and operated by the Skynet Robotic Telescope Network. 

\section*{Data availability}
The data underlying this article are available in the article and in its online supplementary material.

\bsp
\bibliographystyle{mnras}
\bibliography{gwliblongp}

\label{lastpage}

\end{document}